\documentclass[5p]{elsarticle}

\usepackage{hyperref}
\providecommand{\gsponsor}[3]{\href{#3}{#2}}
\providecommand{\grantnum}[2]{#2}
\usepackage{hyperref}
\providecommand{\linkref}[2][]{%
  \if\relax\detokenize{#1}\relax
    #2%
  \else
    #1~#2%
  \fi
}
\usepackage{booktabs} 
\providecommand{\colrule}{\midrule}
\providecommand{\botrule}{\bottomrule}

\usepackage{amssymb}
\usepackage{amsmath}

\newcommand{\MeV}{\,\mathrm{MeV}}
\newcommand{\GeV}{\,\mathrm{GeV}}
\newcommand{\TeV}{\,\mathrm{TeV}}

\journal{Arxiv gen-ph}

\begin{document}

\begin{frontmatter}

\title{New  sum rules of the Koide type}

\author[orcid={0000-0002-6689-214X}]{{Alejandro} {Rivero}}
\ead{arivero@unizar.es}

\address{organization={Institute for Biocomputation and Physics of Complex Systems (BIFI)}, organization={Universidad de Zaragoza}, city={Zaragoza}, postcode={50018}, country={Spain}}


\begin{abstract}
We report a  mass rule of Koide type with inverse
shape, \[m_i=M^{(d)} (w_0+w_i)^{-2}.\] It applies
to the down-quark sector with
numerical precision comparable to that of the direct
charged-lepton sum rule $m_i=M^{(l)} (z_0+z_i)^{2}$.
For central mass values, Koide ratio reaches exactly $2/3$
near $Q \simeq 280\TeV$ under Standard Model renormalisation-group running.
We also review other rules of the direct kind involving quarks.
\end{abstract}


\end{frontmatter}

\section{Introduction}\label{Xsec1-1}\label{introduction}

The preon-program era, including  its attempts to account for family textures and CKM angles, produced a number of striking mass relations. Among these, one of the most notable survivors is the Koide formula for charged leptons \cite{Koide:1983}
\begin{align}
 \label{charges}
    m_i = M^{(l)} (z_0+z_i)^2,\, \sum_{i=1}^3 z_i=0,\, \frac 13 \sum_{i=1}^3 z_i^2 = z_0^2
\end{align}

\noindent that is exact for lepton pole masses: given as input the
masses of electron and muon, it predicts the tau mass within
experimental precision. The quantities $z_i$ were assumed to
be the abelian charge of some preon. We refer to the two
restrictions as ``averaging constraints'', but it could also be
trace conditions on some matrix realization.

Once the preonic justification disappears, it is not
easy to find a reason for a Yukawa coupling going as a square,
not to say as an additive charge. It is possible to accommodate the relation through
a purpose-built Higgs potential, at the cost of introducing non-standard anomalous
dimensions for the Yukawa operators~\cite{k3}. But the formula anyway was thought
to fail for quarks. This is not quite so; besides the known
mixed tuples we will discuss later, the main goal of this
paper is to give notice of the discovery of an inverse tuple for the
down quarks
\begin{equation}
 \label{chargesInv}
    m_i^{(d)} = M^{(d)}/(w_0+w_i)^2,\, \sum w_i=0,\, \frac 13 \sum w_i^2 = w_0^2
\end{equation}
that is exact at some point above 100 TeV and remains within one standard deviation of the experimental central values throughout the running.
In some sense this is the opposite exactness of the lepton tuple,
that keeps stably worse under running but works fine in the\break infrared.

\section{The ratio form}\label{Xsec2-2}

It is conventional to rewrite the Koide relation in a form that eliminates the overall mass scale $M$.
This practice has the advantage
of condensing (\ref{charges}) into a single equation
and the disadvantage of hiding the original additive
charges that motivate the expression. With this format
we would write
\begin{align}
\label{lepton}
K(e,\mu,\tau)=\frac{m_e + m_\mu + m_\tau}{(\sqrt{m_e}+\sqrt{m_\mu}+\sqrt{m_\tau})^2}
     = \frac 13 (1+\frac{\langle z_i^2\rangle}{z_0^2})
     =\frac{2}{3}
\end{align}

for the charged leptons, and
\begin{align}
\label{down}
K^{-1}(d,s,b)=\frac{ {1/m_d} + {1}/{m_s} + {1}/{m_b} }
     { \left({1}/{\sqrt{m_d}}+{1}/{\sqrt{m_s}}+{1}/{\sqrt{m_b}}\right)^2 }
     = \frac{2}{3}
\end{align}

\noindent for the down quarks.

This form is well-defined, provided all charges are positive, as seems to be the case for both tuples.

The averaging constraint $\langle z_i^2 \rangle = z_0^2$
(or respectively with $w$ charges in the inverse formula), $i \in \{1,2,3\}$, is equivalent to the ratio being $2/3$.
With only positive charges, the left-hand side is bounded between $1/3$ and $1$.

In some cases, instead of operating with the square root of
each mass, it can be preferable to define explicitly quantities
$q_i= M^{1/2} (z_0 + z_i)$, where $M$ is the overall mass scale
of the tuple. The connection is then $q_i^2 = m_i$, and one
must take into account that the same particle mass
can be represented by charges of opposite sign
in different tuples.

\section{Other quark tuples}\label{Xsec3-3}

It has been conjectured that the whole of the quark
sector could be covered by generation-alternating tuples
\begin{equation}
(tbc) \qquad (bcs) \qquad  (csu) \qquad (sud)
\label{Xeqn5-5}
\end{equation}
that would correspond to four of the eight possible choices in
\begin{equation}
(ts) \times (ub) \times (cd)
\label{Xeqn6-6}
\end{equation}
Neither tuple fares as well as \linkref[Eqs.]{(\ref{lepton})}, (\ref{down});
the tuples $(tbc)$ and $(bcs)$ are within some percent in
the infrared, especially when one chooses a ``decoupling'' approach
where each mass is taken as $m(m)$ in some running scheme.
The other two tuples fail in an interesting way: they work
under the replacements $u \to 0$ and $d\to (u+d)$ much
like in ChPT approaches.

Instead of the common scale analysis we used in the inverse
down tuple, here we base the observation of
these tuples in PDG 2025 values \cite{ParticleDataGroup:2024cfk} in $\overline{MS}$ scheme:
$m_s(2\GeV) = 93.5 \pm 0.8$,
$m_c(m_c) = 1273.0 \pm 4.6$,
$m_b(m_b) = 4183 \pm 7$ MeV,
$m_t(m_t) =162.69 \GeV$, as we have
found the puzzling fact that they hold better than
at a common scale, and we wonder if it is related to
some process of decoupling. With this caveat we have
the following table, where $K$ denotes the
Koide ratio $\sum q_i^2 / (\sum q_i)^2$ and
we mark with a minus sign the use of the negative
square root for a charge, i.e. ``-s'' indicates
the use of $q_s = -\sqrt{m_s}$

\begin{tabular}{lll}
\toprule
Tuple  & $K$ & Dev from $2/3$ \\
\colrule
$(t,b,c)$                   & $0.662729$ & $-0.591\%$ \\
$(-s,c,b)$           & $0.674802$ & $+1.220\%$ \\
$(c,s,u)$                    & $0.624058$ & $-6.391\%$ \\
$(c,s,0)$,        & $0.664478$ & $-0.328\%$ \\
$(s,u,d)$              & $0.565959$ & $-15.106\%$ \\
$(s,0,u+d)$ & $0.664831$ & $-0.275\%$ \\
\botrule
\end{tabular}\\

In the table,
0 and $u+d$ signal the above replacements, so that the charge associated
to $u+d$ is $\sqrt{m_u+m_d}$ and the whole tuple $(s,0,u+d)$ can be calculated
from the quotient $m_s/\bar m_{ud} = 27.33$ \cite{ParticleDataGroup:2024cfk},
 using then $\sqrt {2 \bar m_{ud}}$ as the charge.

For comparison, we show also the no-replaced tuples $(c,s,u)$ and $(s,u,d)$
choosing as reference values for $u$ and $d$ the ones of \cite[Sec.~60.5.1]{ParticleDataGroup:2024cfk}, namely $2.20\MeV$ and $4.69\MeV$

The substitution $u \to 0$ is interesting both
algebraically and historically. The first mention of a Koide-like tuple
is the $s d u$ set used
by Harari et al. \cite{HarariHW}, substituting $u \to 0$ in the
lower mass quarks as part of another scheme, based only in symmetry, to produce the Cabibbo angle.
As for the other tuples, the possibility of fitting $(t,b,c)$ was first mentioned by
Rodejohann and Zhang \cite[v1]{RodejohannZhang}, and the
tuple $(b,c,s)$ is also well known \cite{Rivero1111}, with the detail of being the only one that requires a negative charge, conventionally assigned to the $s$ quark.

The exact algebraic chain, starting from an exact seed $(0,m_s,m_c)$
instead of experimental values, and generating new elements via
the equation, is an interesting object by
itself. For instance, the inverse tuple also ``sees'' the top quark.
More precisely, the inverse-seesaw solution satisfies
\begin{equation}
m_d\cdot \sigma\!\left(\frac{m_t}{m_c}\right)=m_b,
\label{Xeqn7-7}
\end{equation}
where $\sigma$ denotes the Galois conjugation
$\sqrt{3}\mapsto -\sqrt{3}$. When going across the field
algebra, we find the numbers $(151,28)$ appearing both
in $\frac{m_t}{m_c}=151-28\sqrt{3}$
and also in the calculation of the inverse solution for $m_d$.

Considered as a whole, and adjoining the inverse tuple,
it is amusing that a single formula provides
enough mass relations to adjust the six quarks up to one overall
scale. However, this observation should be interpreted cautiously: as we
relax the precision of the mass rule, it is really not so
complicated to find random masses fitting into them, as far as
they are taken from a log-uniform distribution.

\section{Renormalisation group}\label{Xsec4-4}

\begin{figure}[!t]

\centerline{\includegraphics[width=\columnwidth]{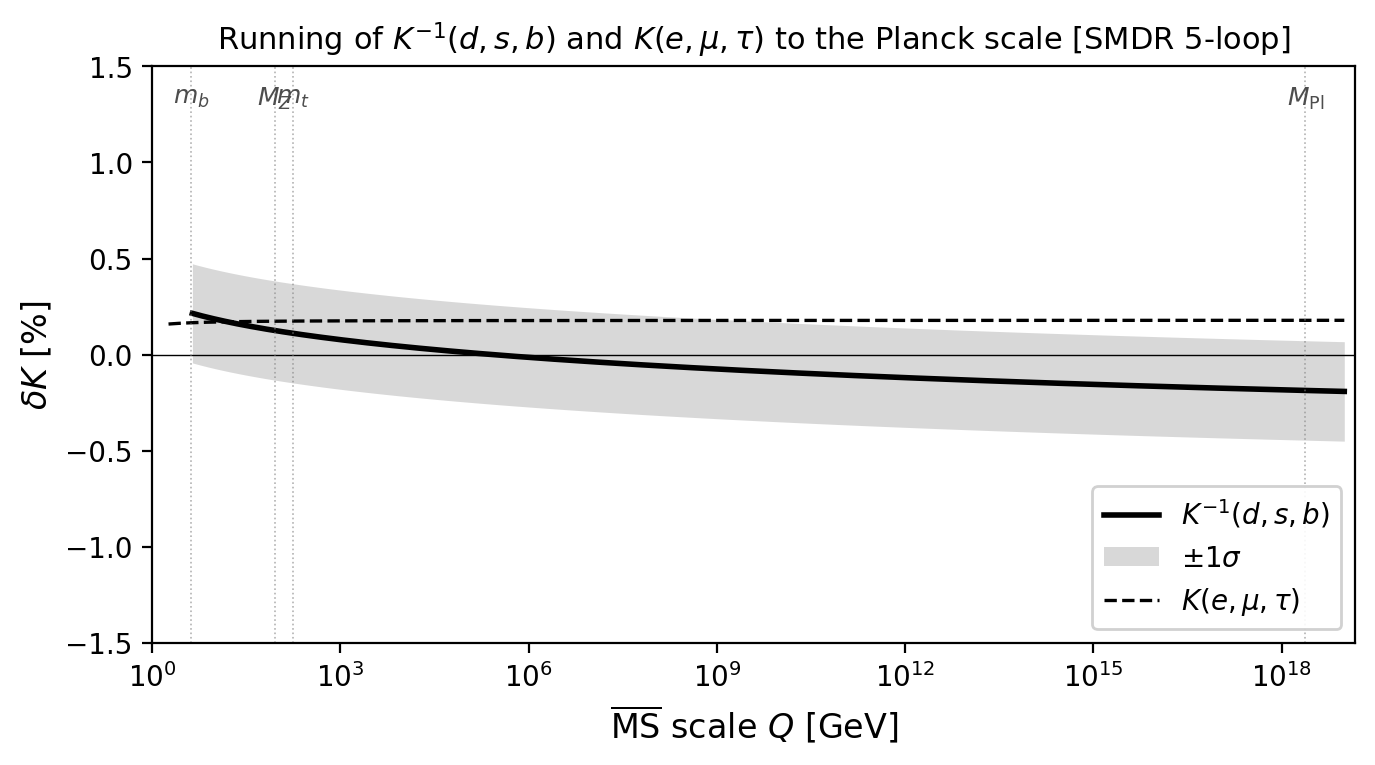}}
\caption{RG running up to Planck mass.}
\label{fig_rg0}
\end{figure}

Quotient tuples between particles of the same
kind have the same exponents under the renormalisation
group, and they are approximately invariant;
the main source of perturbative change is
the Yukawa coupling; that means that low-mass
quark tuples such as $(sud)$ or their variants
run practically flat.

As for the inverse tuple of down-type quarks,
for the desert hypothesis
under renormalisation-group running (5-loop Standard Model,
using the Martin--Robertson SMDR code~\cite{MartinRobertson:2019}),
we can see how the ``inverse Koide ratio'' improves toward the UV:

\begin{tabular}{lll}
\toprule
Scale $Q$ & $K^{-1}(d,s,b)$ & Deviation from $2/3$ \\
\colrule
$M_Z$        & $0.66750$ & $+0.13\%$ \\
$1\TeV$      & $0.66719$ & $+0.08\%$ \\
$100\TeV$    & $0.66675$ & $+0.01\%$ \\
$280\TeV$    & $\mathbf{0.66667}$ & $\mathbf{0.00\%}$ \\
$10^6\GeV$   & $0.66657$ & $-0.01\%$ \\
$M_\text{Planck}$ & $0.66539$ & $-0.19\%$ \\
\botrule
\end{tabular}\\

It crosses $2/3$ at $Q = 280\TeV$ using the default
model of SMDR. It crosses 2/3 a bit earlier,
at $10.18 \TeV$, if we interpolate the 2024-PDG
Table 2 values from \cite{AHS:2024}, which
uses the REAP Mathematica code (permitting simultaneous CKM running) with Yukawa inputs at one digit less precision than our SMDR runs. As we
see in \linkref[Fig.]{\ref{fig_rg0}}, the equation
is within one-sigma error along all the run, so
more experimental precision is needed if we
want to pinpoint the exact scale of a portal of new physics.
The determination of the crossing point is very
sensitive to variations in the mass of the down and
strange quarks. A preliminary run with updated 2025 inputs shifts the crossing slightly towards lower energy but keeps the curve within the one-sigma band throughout the run, so the existence of a crossing is robust against current measurement updates

\section{Other observations}\label{Xsec5-5}

It is interesting to examine the global mass scale of each
tuple,
\begin{equation}
    \label{global}
    M z_0^2 = (m_1 + m_2 + m_3)/6
\end{equation}

The exact chain $(0,a,b) \leftrightarrow (-a,b,c)$
presents an obvious factor $3$ between the corresponding
$M z_0^2$ mass scales. This factor is observed between
$(e,\mu,\tau)$ and $(-s,c,b)$, but not in
the quark chain. With PDG 2025 data
as above, what we observe is a factor $2$. More
precisely, we get
\begin{equation}
\label{factor2}
    z_0^{(-s,c,b)}/z_0^{(0,s,c)} = 1.99974
\end{equation}
with a deviation of 0.013\% , still within
$0.1\sigma$ of the experimental data, if we extract the
$z_0$ from the square roots. On the other hand, if we extract the
quotient from \linkref[Eq.]{\eqref{global}}, the discrepancy with
a factor 2 is greater, we get $2.01522$. This can be interpreted
as a hint that the charges, not their squares, are
the main objects to establish sum rules here.

Another consequence of \linkref[Eq.]{\eqref{factor2}} is
that it implies a sum rule between
charges:
\begin{equation}
    \sqrt{m_b}= 3\sqrt{m_s} + \sqrt{m_c}
\label{Xeqn10-10}
\end{equation}
Which alone predicts $m_b = 4184.5$ MeV vs experimental $4183 \pm 7$, a match to 0.04\% that is noticeable.

While this match is striking, the relation belongs to a broader family $\sqrt{m_3} = (k+1)\sqrt{m_1} + (k-1)\sqrt{m_2}$, and its statistical significance relative to chance requires further study. Generically,
relations between quotients $\sqrt {m_1/m_2}$ were
deeply explored in 1980{--}1990 as source of textures
for the CKM matrix \cite{Fritzsch:1977,Wilczek:1978xi,HarariHW, Fritzsch:1979zq}.

\section{Final discussion}\label{Xsec6-6}

The main conclusion of this note is that out of four
equal charge tuples in the standard
model, namely charged and neutral leptons,
and up and down quarks, two of them are known to
have mass sum rules of the Koide type up to one-sigma
experimental precision. The neutrino tuple cannot
be decided empirically (but see \cite{brannen}),
and the up-type quark seems to need an interleaving
structure to match in this pattern. That the up-type
quarks get a more complicate structure can be seen
in models of flavour. As an example, in the
susy-guided model of \cite{sBootstrap}, when the
flavour-family group is broken to a SU(3) family, the
down type squarks remain in triplets but the up type
squarks seem to need sextets.

That the shape of both sum rules differs by an inversion
adds interest to the intuition, to be examined elsewhere, of obtaining Koide formula in scenarios of duality, such as the ones of Sannino \cite{CacciapagliaSannino} or more generic ones
with explicit link between mesinos and leptons, as in
\cite{sBootstrap}.

A partial support for such Sannino or Seiberg-like
duality scenarios is that it is easy to select actual mesons whose
masses are near those of lepton and quark tuples,
and so reproduce Koide formula in the meson sector. In
some sense this exploits the proximity between
the QCD scale and the (Yukawa-scaled) Fermi scale. Once
the mass of pion is considered as a substitute
of the $\mu$ lepton or the $s$ quark then one can
scan for either the pattern $(0, \pi, D)$ or the pattern $(\pi, D, B)$,
respectively.

There is no obvious clue of the mechanism that
generates the mass as the square of an abelian charge.
The initial mechanism of preons is discouraged by
the existence of a sequence of tuples so that
the same particle would need different decompositions
in each tuple.

The possibility of chaining tuples to produce
new masses is a mildly interesting mathematical
problem involving sequences of
involutions on the quadric equation,
\begin{equation}
 x^2 + y^2 + z ^2 = 4 (xy+yz+zx)
\label{Xeqn11-11}
\end{equation}
while maintaining
agnosticism about the dynamical model of the Koide equation.
It supplements the approach of presenting each tuple
as a rotation of the hypercharge component of a SU(3)
symmetry.
\vfill\pagebreak
\section*{Declaration of generative AI and AI-assisted technologies in the manuscript preparation process}

During the preparation of this work the
author used Claude (Anthropic) and GPT 5.4 (OpenAI)
for algebraic exploration, numerical checks, and minor drafting assistance. The author reviewed and
edited all the content and takes full responsibility for the publication.

The relation motivating this letter was noticed
during an AI-assisted study for a different paper,
namely on the use of Seiberg-like SUSY models to
explain the lepton tuple. It became apparent after
the accidental launch of a numerical exploration of
inverse tuples by Claude Opus 4.6.

\section*{Data availability}
No data was used for the research described in the article.

\section*{Declaration of competing interest}
The authors declare that they have no known competing financial interests or personal relationships that could have appeared to influence the work reported in this paper.

\section*{Acknowledgement}
The author thanks M.~Porter for valuable discussions.

This work was partially supported by Ministerio de Ciencia, Innovación y Universidades (Spain)
and by the \gsponsor{GS501100008530}{European Regional Development Fund}{https://doi.org/10.13039/501100008530} 

(MCIU/AEI/10.13039/501100011033/FEDER, UE)

through grant No. \grantnum{GS501100008530}{PID2022-136374NB-C22}.

\begin{figure}[!t]

\centerline{\includegraphics[width=\columnwidth]{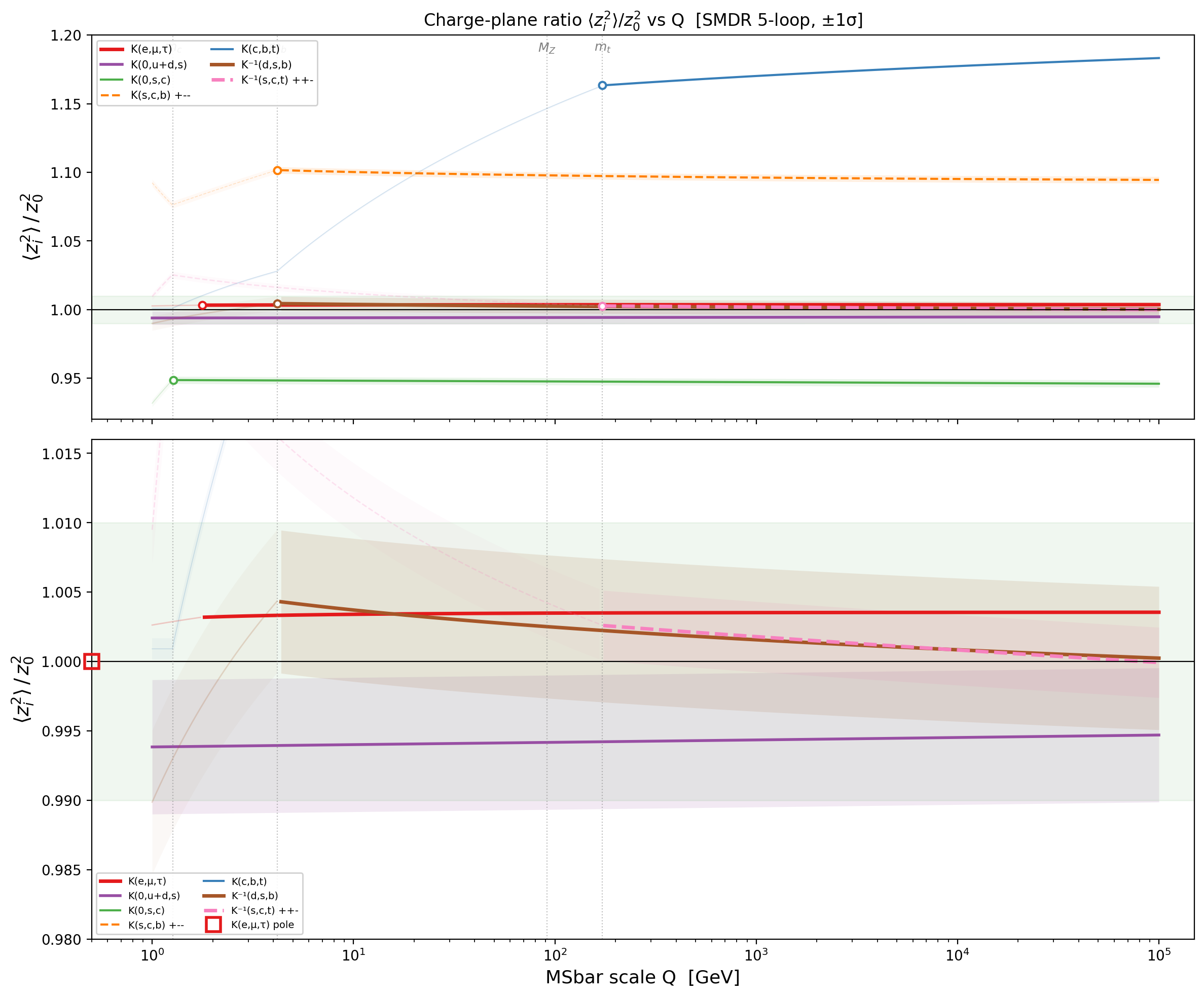}}
\caption{RG running of other koide type sum rules.}
\label{fig_rgPlus}
\end{figure}

\end{document}